\documentclass[journal,10pt]{IEEEtran} 

\usepackage{amsthm}
\usepackage{amsfonts}
\usepackage{amssymb}
\usepackage{mathrsfs}
\usepackage{mathtools}
\usepackage{float}
\usepackage[pdftex]{graphicx}
\usepackage{amsmath}
\usepackage{color}
\usepackage{multirow,multicol}
\usepackage{graphicx}
\usepackage{times}
\usepackage{textcomp}
\usepackage{verbatim}
\usepackage[table]{xcolor}
\usepackage{balance}
\usepackage{lipsum}
\usepackage[inline]{enumitem}
\usepackage{cuted}
\usepackage[caption=false,font=footnotesize]{subfig}
\usepackage{cite}
\usepackage{algpseudocode,algorithm}
\usepackage{hyperref}
\usepackage{tcolorbox}
\usepackage{setspace}

\usepackage[table]{xcolor}
\definecolor{color1}{RGB}{199,209,232}
\definecolor{color2}{RGB}{230,231,233}

\DeclareMathOperator*{\minimize}{minimize\;} 

\makeatletter
\newcommand{\smallsym}[2]{#1{\mathpalette\make@small@sym{#2}}}
\newcommand{\make@small@sym}[2]{%
	\vcenter{\hbox{$\m@th\downgrade@style#1#2$}}%
}
\newcommand{\downgrade@style}[1]{%
	\ifx#1\displaystyle\scriptstyle\else
	\ifx#1\textstyle\scriptstyle\else
	\scriptscriptstyle
	\fi\fi
}
\makeatother

\begin{document}
	
	\title{ Index Modulation for Integrated Sensing and Communications:
		A Signal Processing Perspective}

	\author{\IEEEauthorblockN{ \normalsize
			Ahmet M. Elbir, \textit{Senior Member, IEEE}, Abdulkadir Celik, \textit{Senior Member, IEEE}, \\  Ahmed M. Eltawil, \textit{Senior Member, IEEE}, and Moeness G. Amin, \textit{Fellow, IEEE}}
		\thanks{A. M. Elbir is with the King Abdullah University of Science and Technology, Saudi Arabia; and Duzce University, Duzce, Turkey  (e-mail: ahmetmelbir@ieee.org).}
		\thanks{A. {C}elik and A. M. Eltawil are with King Abdullah University of Science and Technology, Saudi Arabia (e-mail:  abdulkadir.celik@kaust.edu.sa, ahmed.eltawil@kaust.edu.sa).}
		\thanks{M. G. Amin is with Villanova University, Villanova, PA, USA (e-mail: moeness.amin@villanova.edu).}
		
	}
	\markboth{IEEE Signal Processing Magazine} {IEEE Signal Processing Magazine}
	\maketitle

	\begin{abstract}
		A joint design of both sensing and communication can lead to substantial enhancement for both subsystems in terms of size, cost as well as spectrum and hardware efficiency.	In the last decade, integrated sensing and communications (ISAC) has emerged as a means to efficiently utilize the spectrum on a single and shared hardware platform. Recent studies focused on developing multi-function approaches to share the spectrum between radar sensing and communications. Index modulation (IM) is one particular approach to incorporate information-bearing communication symbols into the emitted radar waveforms. While IM has been well investigated in communications-only systems, the implementation adoption of IM concept in ISAC has recently attracted researchers to achieve improved energy/spectral efficiency while maintaining satisfactory radar sensing performance. This article focuses on recent studies on IM-ISAC, and presents in detail the analytical background and relevance of the major IM-ISAC applications.
	\end{abstract}

	
	\section{Introduction}
	
	{
		As the demand for ubiquitous wireless connectivity continues to soar, the fifth generation (5G) and beyond wireless networks explore new ways to efficiently utilize the scarce wireless spectrum and reduce escalating hardware costs. A prominent strategy in this endeavor is the integration of sensing and communications (ISAC) paradigms to share spectrum between radar and communications subsystems on a dual-functional hardware platform~\cite{jrc_SPM_Moeness_Hassanien2019Sep,mishra2019toward}.  ISAC designs follow two directions: radar-communications coexistence (RCC)~\cite{mishra2019toward} and dual-functional radar-communications (DFRC)~\cite{elbir2021JointRadarComm}. {Specifically, RCC involves both subsystems with their waveforms, which account for the inter-subsystem interference. In contrast, DFRC aims performing both sensing and communication (S\&C) tasks simultaneously via employing a common waveform.} Lately, ISAC has been widely studied as a potent means to efficiently utilize spectrum and thereby save cost	and power via employing a shared hardware platform with common signaling mechanism optimized for S\&C tasks~\cite{elbir2021JointRadarComm}.  Nevertheless, the deployment of ISAC methods typically entails performance trade-offs in  S\&C due to the exploitation of common resources such as spectrum, power, antennas, and waveform optimizations~\cite{heath2016overview,elbir2022Nov_Beamforming_SPM}. As a result, the ongoing research in this area is concentrated on developing spectrum sharing techniques aimed at delivering high communication performance while maintaining satisfactory radar sensing capabilities. A novel approach in this context is index modulation (IM), which refers to using radar resources, specifically the radar waveforms,  to index the information-bearing signals into the radar waveforms, thereby enriching the utilization of the shared spectrum~\cite{im_signalling_ISAC_Hassanien2016Oct, jrc_SPM_Moeness_Hassanien2019Sep}.

		Over the last decade, IM has gained prominence, mainly and merely for communications applications, to achieve superior energy/spectral efficiency (EE/SE) compared to conventional modulation schemes in both long-term evolution (LTE)~\cite{indexMod_Survey_Mao2018Jul} and the 5G new radio (NR) systems, especially those incorporating massive multiple-input multiple-output (MIMO) configurations~\cite{im_subcarrier_Abu,indexMod_Survey_Mao2018Jul,im_ComMag_Wang2021Apr}. IM transmitters uniquely encode additional information into the indices of various system parameters such as subcarriers~\cite{indexMod_Survey_Mao2018Jul}, antennas~\cite{antenna_grouping_SM,jrc_spim_sm_Ma2021Feb}, spatial paths	~\cite{spim_bounds_JSTSP_Wang2019May,spim_BIM_TVT_Ding2018Mar,spim_GBM_Gao2019Jul,spim_onGSM_He2017Sep,spim_lowComplexGSM_Shi2021Jan,spim_GBMM_Guo2019Jul,im_ISAC_bookChapterBibEntry}, etc. {In the literature, the implementation of IM over spatial domain (e.g., location of the activated antennas and/or the pairing among the spatial paths) is called spatial modulation (SM)~\cite{jrc_spim_sm_Ma2021Feb,spim_bounds_JSTSP_Wang2019May}. Furthermore, a more general scenario is regarded as generalized spatial modulation (GSM), which involves the IM of information bits mapped to both constellation symbols as well as spatial symbols~\cite{antenna_grouping_SM,spim_lowComplexGSM_Shi2021Jan}.    }
		
		
		By leveraging indices of aforementioned resource entities and radio-frequency (RF) components (e.g., subcarriers, time slots, channel states, antennas or beamformers), IM-aided systems have demonstrated significant performance improvements in the communications-only systems~\cite{spim_GBMM_Guo2019Jul,spim_GBM_Gao2019Jul,spim_bounds_JSTSP_Wang2019May}. In particular, the IM-aided systems deliver even higher SE than fully digital beamformers in communications-only systems; attributed to the additional information bits encoded via the wide variety of network resources and hardware components~\cite{spim_GBMM_Guo2019Jul,spim_BIM_TVT_Ding2018Mar,spim_GBM_Gao2019Jul}.
		
		Building on the success in communications-only MIMO systems, IM has recently been applied to ISAC to improve performance by compensating for losses caused by resource and hardware sharing~\cite{im_OFDM_ISAC_Huang2021Feb,spim_JRC_FRAC_IM_Ma2021Oct,jrc_generalized_SM_Xu2020Sep,im_majorcom_IndexMod_Huang2020May,im_ISAC_sparseArray_Wang2018Aug}. {In a manner akin to the orthogonal frequency division multiplexing (OFDM)-MIMO systems, IM in OFDM-ISAC assigns subcarrier signals to antennas either  exclusively~\cite{im_OFDM_ISAC_Huang2021Feb} or in a shared fashion~\cite{jrc_generalized_SM_Xu2020Sep,ref_R2_Xu} to improve SE while preserving waveform orthogonality.} The employment of sparse antenna arrays in ISAC provides a means of exploiting IM over the antenna indices~\cite{im_ISAC_sparseArray_Wang2018Aug,elbir_sparseArrayRadar_chapter_CSDOAEst}.  Furthermore, spatial path IM  (SPIM) schemes, devised for both millimeter-wave (mmWave)~\cite{elbir_SPIM_MMWAVE_RadarConf_Elbir2022Nov} and terahertz (THz)~\cite{elbir_SPIM_ISAC_THZ_Elbir2023Mar} ISAC systems, involve the entire antenna array, and a switching network of phase shifters is used to modulate information bits into the indices of spatial paths between the transceivers.

		{Historically, ISAC initiatives have revolved around generating a unified waveform by amalgamating communication symbols with radar signals, which has led to the development of numerous information embedding techniques that integrate or represent information bits into or by radar waveforms~\cite{jrc_SPM_Moeness_Hassanien2019Sep} with continuous phase modulation (CPM) on a frequency agile radar (FAR)~\cite{im_majorcom_IndexMod_Huang2020May}, linear frequency modulation (LFM)~\cite{im_radarPulseModulation_Nowak2016Oct}, frequency modulated continuous waveform (FMCW)~\cite{spim_JRC_FRAC_IM_Ma2021Oct}, frequency-hopping (FH)~\cite{fh_MIMO_radar_Wu2021Dec,im_hybrid_Xu2022Nov} and the foundation of the radar sub-pulses~\cite{im_waveformDesign_radarBasis_Wu2020Oct}.} In~\cite{im_waveformPerm_Hassanien2018Dec}, a waveform shuffling was used to embed information using the association of a waveform with an antenna, which is considered as a form of {SM}. It is worth-mentioning that a special case of IM is code-shift keying (CSK) where a communication symbol is mapped to a radar waveform~\cite{im_FH_Eedara2022Jan}.  These approaches enable the joint systems to concurrently perform radar target detection and communication symbol transmission.

		While the above techniques induce minimal effects on the radar performance,  the unification of a single waveform typically results in inferior communication rates. Therefore, an {SM} approach is pursued in~\cite{jrc_spim_sm_Ma2021Feb}, wherein IM is performed over both antenna and frequency indices in such a way that communication symbols are conveyed in the selection of antenna indices with distinct waveforms for radar and communication tasks. Specifically, the array is bisected into two subarrays: one for transmitting communication symbols and the other for radar signals with chirp waveforms.  A similar SM approach is delineated in~\cite{jrc_generalized_SM_Xu2020Sep}, wherein selected antennas transmit at shared subcarrier waveforms to improve the communication rate unlike the exclusive waveform usage as in~\cite{jrc_spim_sm_Ma2021Feb}. To optimize the selection of antenna and subcarrier combinations,  a constellation randomization pre-scaling (CRPS) is proposed in~\cite{im_prescaling_Chen2022Dec} to achieve an index codebook optimized to yield minimum Euclidean distance. Similarly, the Cram\'er-Rao bounds (CRBs) of the delay and Doppler parameters are optimized in~\cite{im_constellationExtension_Memisoglu2023May} to leverage the  constellation extension of the IM-ISAC system.
		
		Another IM approach is changing the transmitter array weights depending on the symbol emitted by using radar parameters. Sidelobe modulation under the auspices of DFRC systems has been developed for radar-centric platforms where the array coefficients are altered for every combination of user symbols~\cite{im_beampatternMod_Hassanien2015Dec}. The same approach underlies directional modulation (DM), a widely used technique in secure communications~\cite{im_directional_Daly2009Jul,im_directional2_Ding2014Feb}.  Both of these methods are viewed as SM~\cite{jrc_SPM_Moeness_Hassanien2019Sep,im_signalling_ISAC_Hassanien2016Oct}.
		
		To improve the communication rate of the IM-ISAC systems, {hybrid index modulation} (HIM) approaches are devised in~\cite{spim_JRC_FRAC_IM_Ma2021Oct} and \cite{im_hybrid_Xu2022Nov} to enhance the {degree-of-freedom (DoF)} by utilizing multiple transmission media, i.e., frequency, phase, and the antenna index of the integrated waveform. Specifically, FMCW-based signaling is suggested to transmit a portion of the information bits via IM over antenna indices by selecting multiple waveforms with distinct frequencies while the remaining bits are transmitted via phase modulation (PM).	To further improve the DoF, an FH signaling strategy for MIMO (FH-MIMO) radar systems is considered in~\cite{im_hybrid_Xu2022Nov}, wherein the communication symbols are embedded into the index of three-tuples encompassing diversity elements of frequency code, the initial phase, and the transmit antenna index. By exploiting the high array gain of the large arrays and the high sparsity of mmWave/THz channels, a  communication-centric modulation scheme, a.k.a. {delay alignment modulation} (DAM), is put forth in~\cite{im_delayAlignmentMod_Xiao2023Apr} to improve the communication rate.	
		

		The primary limitation of the aforementioned IM schemes is that radar sensing is often prioritized, with communication symbols typically embedded into radar waveforms such as CPM~\cite{im_majorcom_IndexMod_Huang2020May}, LFM~\cite{im_radarPulseModulation_Nowak2016Oct}, FMCW~\cite{spim_JRC_FRAC_IM_Ma2021Oct}, FH~\cite{im_hybrid_Xu2022Nov}. Therefore, these approaches tend to yield suboptimal communication rates since the information embedding is usually applied on the slow-time scale across the radar pulses, inevitably bounding the data rate by the radar pulse repetition frequency (PRF). This motivates the development of more communication-centric design to enhance the communication data rate while maintaining radar functionalities. In contrast, communication-centric IM-ISAC designs based on SPIM~\cite{elbir_SPIM_ISAC_THZ_Elbir2023Mar,elbir_SPIM_MMWAVE_RadarConf_Elbir2022Nov} and FH-IM~\cite{im_hybrid_Xu2022Nov,fh_MIMO_radar_Wu2021Dec}, have shown to provide much higher communication rate than that of aforementioned signaling at the cost of complexity and hardware for transceiver design.
		
		

		This article presents an overview of the aforementioned approaches while encapsulating the major	advancements in the information embedding techniques for IM-ISAC, also shedding light on nascent	application arenas. Specifically, the article has a special emphasis on: 1) outlining the main developments in IM for communications-only (briefly) and ISAC systems; 2) presenting the relevant signal processing tools for IM-ISAC; and 3) reflecting on the major challenges and current opportunities for the signal processing research in IM-ISAC.

		\textit{Notation:} Throughout this paper, uppercase and lowercase bold letters denote matrices and vectors, respectively. The conjugate, transpose, and conjugate transpose operations are denoted by $(\cdot)^*$, $(\cdot)^\textsf{T}$ and $(\cdot)^{\textsf{H}}$, respectively. For a matrix $\mathbf{A}$ and a vector $\mathbf{a}$; $[\mathbf{A}]_{ij}$ and $[\mathbf{a}]_k$ correspond to the $(i,j)$-th entry of $\mathbf{A}$ and the $k$-th entry of $\mathbf{A}$, while $\angle\{a\}$ computes the phase of $a$. $\mathbf{I}_N$ is the identity matrix of size $N\times N$,  $\mathrm{vec}\{\cdot\}$ is the vectorization operator, and $\otimes$ denotes the Kronecker product. $\|\mathbf{a} \|_2 = (\sum_{i = 1}^{N} |a_i |^2)^{\frac{1}{2}}  $  denotes the $l_2$-norm,  $\lfloor \cdot \rfloor$ stands for the nearest integers towards minus infinity, and finally, $C(k,n) $ represents the Binomial coefficient for selecting $k$ elements from $n$ as $C(k,n) =  {\footnotesize\left(\begin{array}{c}
				n \\
				k
			\end{array}  \right) } = \frac{n!}{k! (n-k)!} $.

		\begin{figure*}
			\centering
			{ \includegraphics[draft=false,width=.9\textwidth]{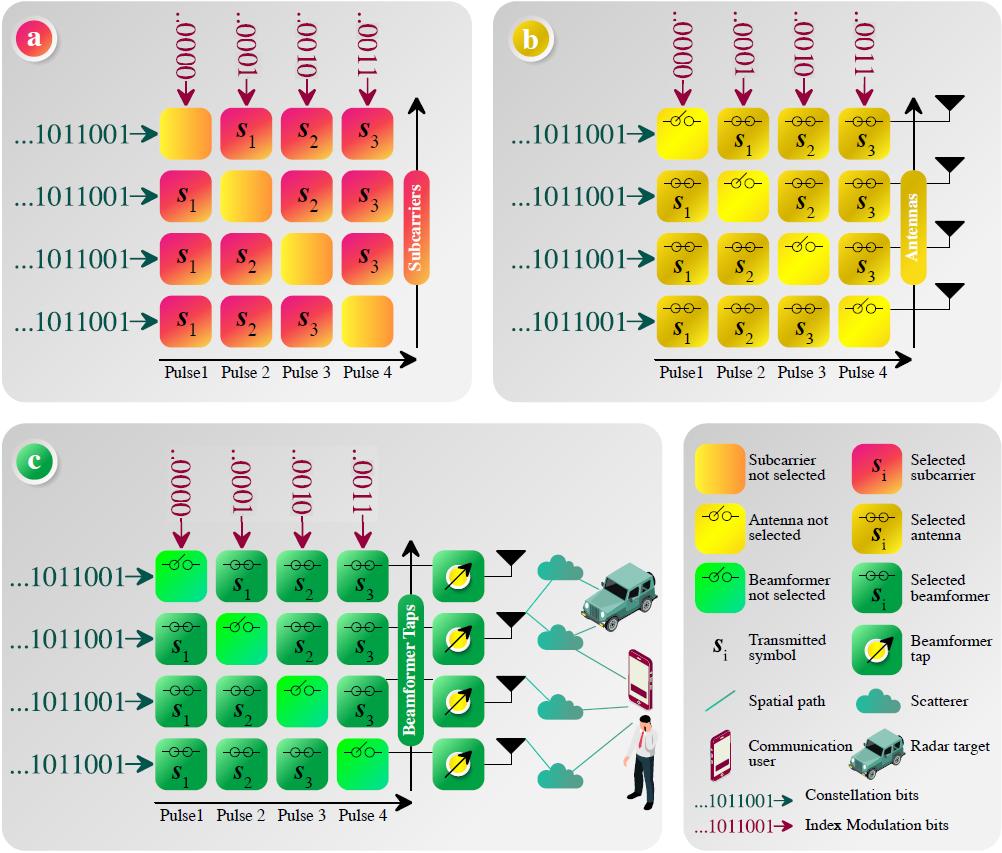} } 
			\caption{IM over different domains, i.e., (a) subcarriers, (b) antennas, and (c) spatial path indices for ISAC. For the IM over subcarrier and antenna indices, a subset of subcarriers ($K_s$ out of $K$) and antennas ($N_s$ out of $N$) are (de)activated, respectively. 
				For spatial path IM, the beamformer taps correspond to the (de)activated spatial paths ($L_s$ out of $L$). In each pulse interval, the information bits are split into two groups: IM bits and constellation bits. The IM bits are utilized by selecting the indices of the IM domain (i.e., subcarrier, antenna and beamformer taps) while the remaining bits are  transmitted via digital constellation.  }
			\label{fig_IM_techniques}
		\end{figure*}
	}
	
	\section{Preliminaries on IM Techniques}
	The fundamental concept of IM capitalizes on the diversity inherent in various system parameters and hardware components (e.g., subcarrier frequency, time slots, channel states, antennas or beamformers).  By exploiting  the indices of these entities, IM facilitates the modulation of information bits, which are subsequently accurately demodulated at the receiver. In this framework, a portion of the information bits is conveyed via conventional communication system, e.g., through PM with signal constellations, while the remaining portion of bits are transmitted via IM. This dual approach substantially enhances the communication data rate by transmitting more information bits concurrently. Herein, we briefly introduce prominent examples of IM techniques, which are illustrated in Fig.~\ref{fig_IM_techniques}. 
	
	\subsection{IM over Subcarrier Frequencies}
	Consider an OFDM scheme with $K$ subcarriers operating on a frequency-selective Rayleigh fading channel,	wherein $\bar{B}$ information bits are employed for a given OFDM subblock. $\bar{B}$ bits are split into two subgroups of $B_i$ and $B_s$ bits, which are called \textit{index bits} and \textit{symbol bits}, respectively. Then, the implementation of IM to transmit $B_i$ bits is performed via employing only $K_s$ out of $K$ active indices, which can be determined by a selection procedure~\cite{im_subcarrier_Abu,indexMod_Survey_Mao2018Jul}. The remaining $B_s$ bits are then mapped on to the $M$-ary signal constellation to determine the data symbols. Fig.~\ref{fig_IM_techniques}(a) illustrates the transmission of index and symbol bits, such that the allocation of the symbol bits changes across the radar pulses, which provides unique representation to modulate more information bits via IM. As a result, we have  $B_s = K_s \log_2 M$ for symbol bits while the index bits via selecting $K_s$ out of $K$ subcarrier frequency indices yields $B_i = \lfloor \log_2 C(K_s,K)\rfloor $. Note that $K_s = K$ corresponds to conventional transmission scheme without IM, i.e., all subcarriers are activated. In comparison, IM over OFDM subcarriers is capable of embedding additional information bits on the activated subcarriers without extra energy consumption.
	
	\subsection{IM over Antenna Indices}
	In massive MIMO systems with large antenna arrays, hybrid beamforming exhibits an efficient solution for reducing the number of RF chains~\cite{elbir2021JointRadarComm}. However, the number of independent data streams is limited by the number of RF chains, thereby restricting the potential spatial multiplexing gain. Performing IM over the large number of antennas can compensate for this limitation and provide substantial improvements in SE~\cite{antenna_grouping_SM,im_antennas_generalized_Zhang2013Oct}. This is achieved by transmitting additional information bits via the indices of the antennas in the array as illustrated in Fig.~\ref{fig_IM_techniques}(b). Let $N$ denote the number of antennas in the array, then activating $N_s$ out of $N$ antennas with IM yields $\lfloor \log_2 C(N_s,N)\rfloor$ possible choices for \textit{antenna selection pattern}~\cite{jrc_SPM_Moeness_Hassanien2019Sep}. 
	In practical implementation, 	the symbol bits are modulated by $M$-ary constellation mapper, whereas the index bits are fed to an antenna index selector to activate $N_s$ antennas. As a result,  IM over antenna indices provides a communication data rate of $\lfloor \log_2 C(N_s,N)\rfloor + \log_2 M$, compared to conventional transmission which only achieves $\log_2 M$ {per RF chain}.
	
	\subsection{IM over Spatial Path Indices}
	In SPIM, the diversity of the scattering paths between the transmitter and the receiver is exploited to convey additional information bits. Fig.~\ref{fig_IM_techniques}(c) presents the implementation of SPIM, wherein each spatial path is represented by a beamformer, which is realized by analog phase shifters.	Therefore, a switching mechanism between the RF chains and the beamformer taps, each of which corresponds to a selected spatial path, establishes a \textit{spatial pattern}. Let $L$ denotes the number of spatial paths, $L_s$ out of $L$ paths are used for IM, which yields $C(L_s,L)$ choices of connection to incorporate the spatial domain information~\cite{spim_bounds_JSTSP_Wang2019May,spim_BIM_TVT_Ding2018Mar}.  As a result, there exists $2^{\lfloor \log_2 C(L_s,L)\rfloor}$ spatial patterns, each of which is generated by activating the corresponding analog beamformer tap in the transmitter~\cite{elbir_SPIM_ISAC_THZ_Elbir2023Mar}.

	

	%
	%
	%

	\begin{table*}
		\caption{State-of-the-Art IM-ISAC Techniques Over The Indices of IM Media: (F)requency, (A)ntenna, (P)hase, and (S)patial (P)ath 
		}
		\footnotesize
		\label{tableSummary}
		\centering
		\begin{tabular}{p{0.005\textwidth}p{0.005\textwidth}p{0.005\textwidth}p{0.01\textwidth}p{0.32\textwidth}p{0.18\textwidth}p{0.3\textwidth}}
			\hline 
			\hline
			\multicolumn{4}{c}{ \centering\arraybackslash\bf IM Media} 
			&\centering\arraybackslash\multirow{2}{*}{	\centering\arraybackslash\bf Signal Processing Tools
			}	&\centering\arraybackslash\multirow{2}{*}{	 \centering\arraybackslash\bf Communication Rate
			}&\centering\arraybackslash\multirow{2}{*}{\centering\arraybackslash	 \bf Advantages/Drawbacks
			} \\ 
			\bf F&\bf A& \bf P& \bf SP &
			\\
			\hline
			\cellcolor{color2}	\center\arraybackslash		\bf\checkmark
			\cellcolor{color2}& \bf 
			\cellcolor{color2}&\bf 
			\cellcolor{color2}&\bf 
			\cellcolor{color2}	&\cellcolor{color1}\arraybackslash Allocating $K_s$ out of $K$ subcarriers with null subcarrier detection and Golay complementary sequences, and maximum likelihood (ML) estimation for decoding~\cite{spim_OFDM_IM_JRC_Sahin2021}
			&\cellcolor{color2}\center\arraybackslash  \vfill
			$\lfloor \log_2 C(K_s,K)  \rfloor$ \vfill
			
			&\cellcolor{color1} \arraybackslash {OFDM subcarriers are disjointly allocated for radar and communications	while null subcarriers reduce communication rate} \\
			\hline 
			\cellcolor{color2}		\center\arraybackslash		\bf\checkmark	& 
			\cellcolor{color2}		&\center\arraybackslash		\bf\checkmark
			\cellcolor{color2}		&
			\cellcolor{color2}		&
			\cellcolor{color2}		\cellcolor{color1}\arraybackslash PM with $M$-ary constellation and $K_s$ out of $K$ subcarriers with ML decoder~\cite{im_ofdm_ISAC_Li2024Feb}
			&\cellcolor{color2}\center\arraybackslash \vfill
			$K_s\log_2 M + \lfloor \log_2 C(K_s,K)  \rfloor$ \vfill
			
			& \cellcolor{color1}\arraybackslash  Improved data rate via PM over symbol bits while exploiting only a single IM domain of subcarriers	 \\
			\hline 
			\cellcolor{color2}		&\center\arraybackslash		\bf\checkmark
			\cellcolor{color2}		& 
			\cellcolor{color2}		&
			\cellcolor{color2}		&
			\cellcolor{color2}		\cellcolor{color1}\arraybackslash Transmission of $N_s$ orthogonal waveforms over $N_s$ out of $N$ antennas, which are optimized via sequential convex programming~\cite{im_ISAC_sparseArray_Wang2018Aug}
			&\cellcolor{color2}\center\arraybackslash  \vfill
			$\lfloor \log_2 C(N_s,N)  \rfloor$ \vfill
			
			& \cellcolor{color1}\arraybackslash   Sparsity is exploited to improve the DoF for IM while reducing the array gain and causing distorted beampattern	 \\
			\hline 
			\cellcolor{color2}			& \center\arraybackslash		\bf\checkmark
			\cellcolor{color2}		&\center\arraybackslash		\bf\checkmark
			\cellcolor{color2}		&
			\cellcolor{color2}		&
			\cellcolor{color2}		\cellcolor{color1}\arraybackslash Allocating $N_s$ out of $N$ antennas fed by $M$-ary constellation bits decoded via compressed sensing~\cite{jrc_spim_sm_Ma2021Feb}
			&\cellcolor{color2}\center\arraybackslash \vfill
			$N_s\log_2 M + \lfloor \log_2 C(N_s,N)\rfloor$\vfill
			
			& \cellcolor{color1}\arraybackslash 	Enhanced communication rate thanks to PM while nonuniformity of the subarrays cause distorted beampattern and reduced array gain \\
			\hline 
			\cellcolor{color2}	\center\arraybackslash		\bf\checkmark	& \center\arraybackslash		\bf\checkmark
			\cellcolor{color2}		&
			\cellcolor{color2}		&
			\cellcolor{color2}	&
			\cellcolor{color1}\arraybackslash Employing $N$ antennas for the transmission of $N$ waveforms selected from $K$ frequencies with orthogonal matching pursuit (OMP)-based decoding~\cite{im_majorcom_IndexMod_Huang2020May}
			&\cellcolor{color2}\center\arraybackslash\vfill
			$\lfloor N \mathrm{log}_2 K \rfloor$ \vfill
			
			&\cellcolor{color1} \arraybackslash 	Exploiting high array gain with full array while limited communication rate due to performing IM over only frequency indices without PM    \\
			\hline 
			\cellcolor{color2}	\center\arraybackslash		\bf\checkmark& \center\arraybackslash		\bf\checkmark
			\cellcolor{color2}		&
			\cellcolor{color2}			&
			\cellcolor{color2}			&
			\cellcolor{color1}\arraybackslash Partitioned $N$-element array with $G$ groups fed by $K$ waveform frequencies with OMP-based decoder~\cite{im_majorcom_IndexMod_Huang2020May}
			&\vfill
			\cellcolor{color2}\center\arraybackslash $\lfloor G\log_2 K \rfloor + \lfloor N\log_2 G \rfloor$\vfill
			
			&\cellcolor{color1} \arraybackslash 	Enhanced array gain with full array of grouped subarrays while limited IM bits are achieved due to IM over frequency and group indices   \\
			\hline 
			\cellcolor{color2}		\center\arraybackslash		\bf\checkmark	& \center\arraybackslash		\bf\checkmark
			\cellcolor{color2}		&\center\arraybackslash		\bf\checkmark
			\cellcolor{color2}		&
			\cellcolor{color2}			&
			\cellcolor{color1}\arraybackslash Transmission of $M$-ary constellation bits, decoded via successive orthogonal decoding, over $N_s$ frequencies through $N_s$ out of $N$ antennas~\cite{spim_JRC_FRAC_IM_Ma2021Oct}
			&\cellcolor{color2}\center\arraybackslash \vfill
			$ {{N_\mathrm{S}\log_2 M} } + \lfloor \log_2 C(N_s,K)\rfloor + \lfloor \log_2 C(N_s,N) \rfloor + \lfloor \log_2 N_s! \rfloor $ \vfill
			
			&\cellcolor{color1} \arraybackslash  Low complexity FMCW signaling with IM over multiple domains, i.e., antenna, frequency and phase indices while achieving more DoF is possible via different  schemes   	 \\
			\hline 
			\cellcolor{color2}	\center\arraybackslash		\bf\checkmark	& \center\arraybackslash		\bf\checkmark
			\cellcolor{color2}		&\center\arraybackslash		\bf\checkmark
			\cellcolor{color2}		&
			\cellcolor{color2}		&
			\cellcolor{color1}\arraybackslash Transmission of $K$ frequency codes, optimized via genetic-algorithms, through $N$ antennas over $H$ subpulses~\cite{im_hybrid_Xu2022Nov}
			&\cellcolor{color2}\center\arraybackslash \vfill
			$H \cdot  (N $ $+ \lfloor \log_2 (C(K,N)\cdot N!) \rfloor  )  $ \vfill
			
			&\cellcolor{color1} \arraybackslash  Exploiting the diversity of multiple resource entities to achieve higher DoF while the transceiver design and demodulation are costly  	 \\
			\hline 
			\cellcolor{color2}			&
			\cellcolor{color2}			&
			\cellcolor{color2}			&
			\cellcolor{color2}		\center\arraybackslash		\bf\checkmark	&
			\cellcolor{color1}\arraybackslash Selecting $L_s$ out of $L_C$ analog beamformers, obtained via alternating optimization, corresponding to each path~\cite{elbir_SPIM_ISAC_THZ_Elbir2023Mar}
			&\cellcolor{color2}\center\arraybackslash\vfill
			$ \lfloor \log_2 C(L_s,L_C) \rfloor   $ \vfill
			
			&\cellcolor{color1} \arraybackslash  Communication-centric design with high data rate while leveraging IM only over the spatial paths \\
			\hline 
			\hline
		\end{tabular}
	\end{table*}

	\section{IM Meets ISAC }
	Unlike the communications-only systems, the ISAC setup requires the joint access to the spectrum, which makes the implementation of IM more challenging since the joint system should account for both S\&C functionalities. 	{From the transmitter perspective, the ISAC transmitter should account for generating multiple beams simultaneously toward both radar target and the communication user for both target detection/tracking and delivering the communication data to the user. This requires collecting both radar-related (e.g., target direction and ranges) and communication-related (e.g., wireless channel matrix) data for the design of joint radar-communication beamforming~\cite{elbir2021JointRadarComm}.  } {From the receiver perspective, the communication receiver is synchronized with the IM-ISAC transmitter to enable demodulation and recovery of the transmitted as well as IM symbols~\cite{im_recovery2_GarciaRodriguez2015May,im_comm_recovery1_Xiao2016Apr,spim_JRC_FRAC_IM_Ma2021Oct,jrc_generalized_SM_Xu2020Sep,im_ISAC_sparseArray_Wang2018Aug,im_radarWaveform_Sahin2017May,im_hybrid_Xu2022Nov}.  Furthermore, the communication receiver should have knowledge of the transmitted ISAC waveforms for successful recovery of the IM symbols.    }

	In the following, we overview the major IM-ISAC schemes and the relevant transmitter architectures. A summary of these techniques is presented in Table~\ref{tableSummary}.


	\subsection{Subcarrier Index Modulation}
	Inspired from the communication-only systems, wherein OFDM waveform is mostly considered, subcarrier IM is applied to ISAC with OFDM signaling. Earliest studies include the use of IEEE 802.11p OFDM signaling, i.e., dedicated short-range communication (DSRC), for vehicular communication and radar sensing by exploiting OFDM packets over subcarriers and time-slots~\cite{ofdm_DFRC_Nguyen}. The implementation of IM with OFDM in ISAC systems is performed by simply activating a portion of the subcarriers during transmission. While (de)activating the subcarriers provides IM for communication, the remaining subcarriers are employed for radar sensing. In particular, for an OFDM-IM systems with  $K$ subcarriers, the transmitted baseband signal is
	\begin{align}
		x_\mathrm{S}(t) = \sum_{b=0}^{B-1} \sum_{{k}} \alpha_{b,{k}} e^{ \mathrm{j} 2\pi f_{{k}} t } \mathrm{rect}\left(\frac{t - bT_0}{T_0}\right), \label{si_xs}
	\end{align}
	where {${k} = 0,\cdots, K_s-1$} denotes the activated subcarrier index selected from the set $\{0,\cdots, K-1 \}$;  $\alpha_{b,k}$ is a complex-valued attenuation factor;  $\mathrm{rect}\left(\frac{t - bT_0}{T_0}\right)$ represents a time-domain rectangular window of duration $T_0$; and $b$ denotes the OFDM symbol index from a total of $B$ symbols. As a result, the communication symbols are embedded into the radar waveforms with IM, yielding the data rate of $\lfloor \log_2 C(K_s,K)\rfloor$. In this setup, a portion of the OFDM subcarriers are allocated for radar while the remaining ones are reserved for communication tasks~\cite{spim_OFDM_IM_JRC_Sahin2021}. Furthermore, a portion of the subcarriers are intentionally left empty. These null subcarriers are distributed intelligently in order to find the location of communication/radar symbols. At the receiver side, the location of the subcarriers associated with radar and communication tasks are found via estimating the location of the null subcarriers. {While this approach is useful in combining  the S\&C functionalities in multi-carrier setup, reserving the subcarriers  exclusively for sensing and having null subcarriers would reduce the communications data rate and limit the radar performance. In essence,  the allocation of the subcarriers for communications is typically determined by the base station prior to data transmission using other criteria not accounting for IM. As such, the possible number of subcarrier configurations available  for IM becomes limited, adversely impacting the achievable rate. } To improve the communication rate while accurately recovering the radar range-velocity profile, various techniques, including modulation symbol cancellation~\cite{im_OFDM_2_Singh,im_ofdm_ISAC_Li2024Feb} and frequency detection demodulation~\cite{im_OFDM_ISAC_Huang2021Feb}, are introduced.


	\begin{figure*}
		\centering
		{\includegraphics[draft=false,width=.99\textwidth]{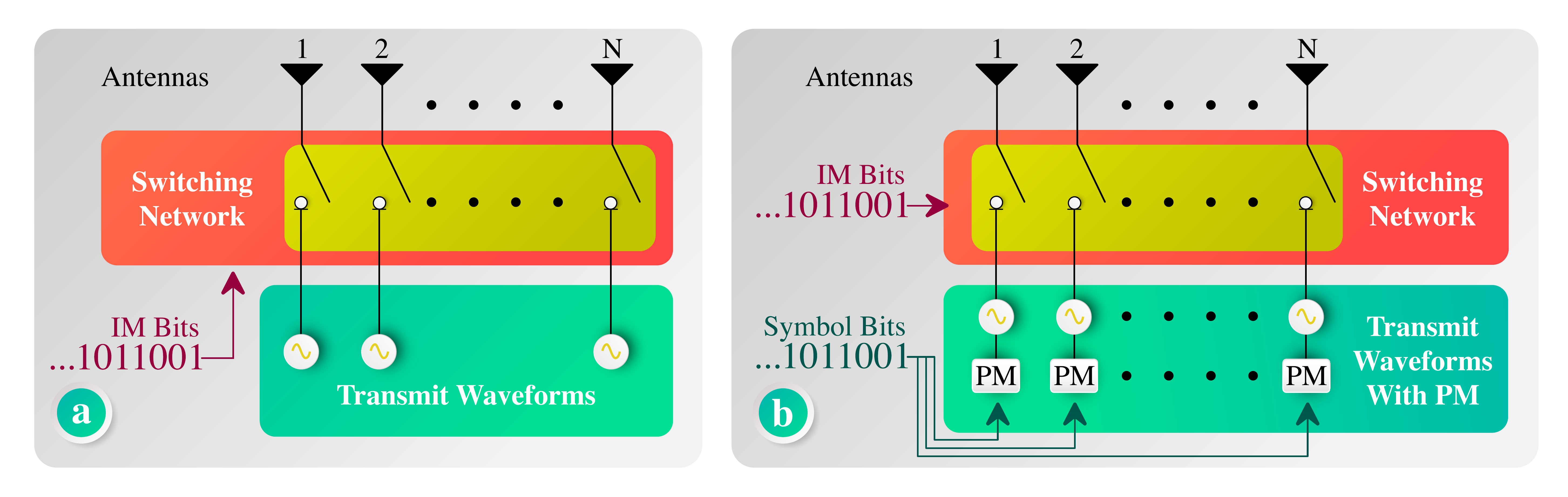} } 
		\caption{IM-ISAC transmitter architectures for IM over antenna indices. In (a), $N_s$ out of $N$ antennas are selected for IM to embed the information bits into the radar waveforms. In (b), a portion of of the information bits are modulated through the selected antenna indices via IM while the remaining symbol bits are phase-modulated with $M$-ary constellation.  }
		\label{fig_IM_ISAC_antenna}
	\end{figure*}

	\subsection{Antenna Index Modulation}
	Unlike employing the full array, sparse arrays provide unique antenna displacements which can be used to embed the communication data symbols into the radar waveforms. In particular, the information bits can be paired with the transmit antenna subarray configurations over different pulse repetition intervals (PRI) without impacting the radar sensing functionality. As depicted in Fig.~\ref{fig_IM_ISAC_antenna}(a), $N_s$ out of $N$ antennas of a MIMO radar are used to form a subarray that transmits $N_s$ orthogonal waveforms $\Psi_n(t)$, $n = 1,\cdots, N_s$, to simultaneously detect the radar targets while sending communication symbols to downlink users~\cite{im_ISAC_sparseArray_Wang2018Aug}. Assume the transmitted signal is echoed from $R$ radar targets with direction-of-arrivals (DOAs) $\{\theta_1,\cdots, \theta_R\}$, and received by an $\bar{N}$-element MIMO receive array, then  the $\bar{N} \times 1$ array output is given by
	\begin{align}
		\mathbf{y} (t, p) = \sum_{r = 1}^{R} \alpha_r(p) \left(\bar{\mathbf{a}}^\textsf{T} (\theta_r,p) \boldsymbol{\Psi}(t)  \right) \tilde{\mathbf{a}}(\theta_r) + \mathbf{n}(t,p), \label{ai_ytp}
	\end{align}
	where $p$ denotes the pulse number; $\alpha_r(p)$ is the reflection coefficient from the $r$-th target at the $p$-th pulse; {$\bar{\mathbf{a}}(\theta_r,p)\in \mathbb{C}^{N_s\times 1}$ and $\tilde{\mathbf{a}}(\theta_r)\in \mathbb{C}^{\bar{N}\times 1}$ represent the steering vectors} of the transmit and receive arrays, respectively; $\boldsymbol{\Psi}(t)\in\mathbb{C}^{N_s}$ includes the transmit radar waveforms as $\boldsymbol{\Psi}(t) = \left[\Psi_1(t), \cdots, \Psi_{N_s}(t)\right]^\textsf{T}$; and $\mathbf{n}(t,p)\in \mathbb{C}^{\bar{N}}$ is a zero-mean vector summarizing the unwanted clutter, interference and noise signals~\cite{im_FH_Eedara2022Jan,im_clutter_radar_Zhang}. In particular, the steering vector for the full array is given by 
	\begin{align}
		{\mathbf{a}}(\theta_r) = \left[1, e^{\mathrm{j}2\pi \frac{{d}}{\lambda} \sin \theta_r },\cdots, e^{\mathrm{j}2\pi\frac{{d}}{\lambda} (N-1)\sin \theta_r}  \right]^\textsf{T},
	\end{align}
	where ${d} = \lambda/2$ denotes the antenna spacing, and  $\lambda = \frac{c_0}{f_c}$ is the wavelength, for which $c_0$ and $f_c$ are the speed of light and the operating frequency, respectively. Denote $\mathbf{Q}(p)$ as the $N_s\times N$ antenna selection matrix as $\mathbf{Q}(p) \in \{0,1\}^{N_s \times N}$. Specifically, for the $(n_1,n_2)$-th element of $\mathbf{Q}(p)$,  $[\mathbf{Q}(p)]_{n_1,n_2} = 1$ represents that the $n_2$-th transmit antenna is chosen for the $n_1$-th sparse array antenna for $n_1\in \{1,\cdots, N_s\}$ and $n_2 \in \{1, \cdots, N\}$. Thus, the $N_s\times 1$ steering vector of the sparse array for the $p$-th pulse is
	\begin{align}
		\bar{\mathbf{a}}(\theta_r, p) = \mathbf{Q}(p) \mathbf{a}(\theta_r).
	\end{align}
	During each radar pulse, a subset of antennas is selected to convey the communication symbols. In particular, $ C(N_s,N)$ different subarray configurations can be generated with a unique selection matrix $\mathbf{Q}(p)$ to embed the communication symbols into the radar waveform. Then, the received signal at a single-antenna communication user of direction $\phi$ is written as
	\begin{align}
		y_C(t,p) = \alpha_C {\mathbf{a}}^\textsf{T} (\phi) \mathbf{Q}^\textsf{T}(p) \boldsymbol{\Psi}(t) + n_C(t,p), 
		\label{userReceived1}
	\end{align}
	where $\alpha_C\in \mathbb{C}$ denotes the channel between the user and the MIMO transmitter while $n_C(t,p)\in \mathbb{C}$ represents the noise term. The received signal in (\ref{userReceived1}) yields the scaled and noisy version of the sparse array steering vector $\bar{\mathbf{a}}(\phi,p)$ after match filtering with $\boldsymbol{\Psi}(t)$ as $	\mathbf{y}_C (p) =  \left[y_{C,1}(p),\cdots, y_{C,N_s}(p)  \right]^\textsf{T}\in \mathbb{C}^{N_s}$, i.e.,
	\begin{align}
		{y}_{C,n} (p) =  \int_T y_C(t,p) \Psi_n^*(t) dt, \hspace{10pt} n =1,\cdots, N_s.
	\end{align}
	Then, the $N_s\times 1$ subarray data is constructed at the user as 
	\begin{align}
		\mathbf{y}_C (p)	= \alpha_C \bar{\mathbf{a}}(\phi,p) + \mathbf{n}_C(p), \label{ai_yc}
	\end{align}
	where $	\mathbf{n}_C (p) =  \left[n_{C,1}(p),\cdots, n_{C,N_s}(p)  \right]^\textsf{T}\in \mathbb{C}^{N_s}$.

	To further improve the communication rate, phase-modulated waveforms are employed in~\cite{jrc_spim_sm_Ma2021Feb}, wherein the selected $N_s$ antennas are used to transmit the symbol bits, which are generated from PM of $M$-ary constellation, as illustrated in Fig.~\ref{fig_IM_ISAC_antenna}(b). As a result, $N_s\log_2 M$ additional symbol bits can be sent.   Then, the number of transmitted bits per symbol becomes $N_s\log_2 M + \lfloor \log_2 C(N_s,N)\rfloor$. The main drawback of the aforementioned IM approaches over sparse arrays is that only a portion of the whole array is utilized, thereby reducing the array gain. In order to exploit the fully array and improve the radar sensing performance, multi-carrier agile joint radar-communication (MAJoRCom) technique is developed in~\cite{im_majorcom_IndexMod_Huang2020May}. While the use of full array does not provide diversity of antenna indices, multiple waveform frequencies are exploited by assigning $K$ distinct waveforms to $N$ antennas. Therefore, the number of transmitted information bits becomes $\lfloor N \mathrm{log}_2 K \rfloor$.


	\subsection{Joint Antenna and Frequency Index Modulation}
\begin{figure*}
		\centering
		{\includegraphics[draft=false,width=.9\textwidth]{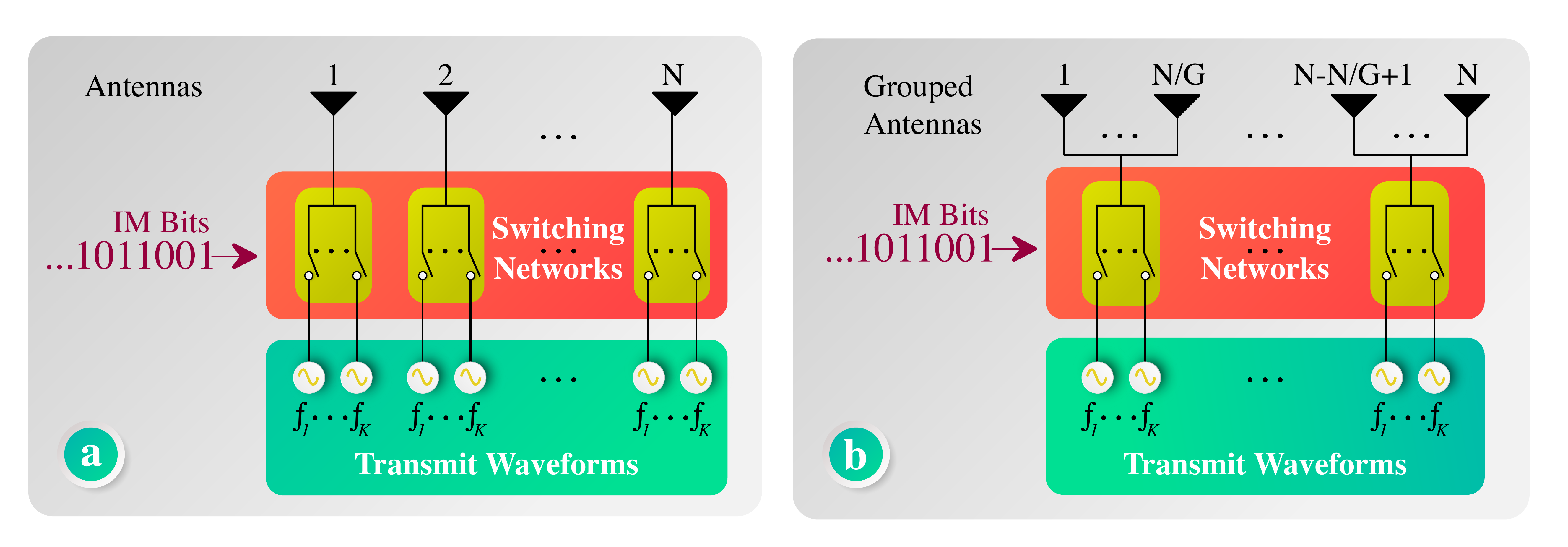} } 
		\caption{Joint antenna and frequency modulation for IM-ISAC. A switching network provides the selection of $N$ out of $K$ waveform frequencies for the transmission of the IM bits via $N$ antennas in (a). To improve the communication rate, the array is partitioned into $G$ groups in (b), each of which are fed from the selection of $N/G$ out of $K$ waveform frequencies.        }
		\label{fig_IM_ISAC_antennaFrequency}
	\end{figure*}

	To further enhance the DoF for IM, multiple resource entities can be combined. One possible way is to employ multiple transmit waveform signals with distinct frequencies~\cite{im_majorcom_IndexMod_Huang2020May}. Suppose that $K$ narrowband signals are transmitted through $N$ antenna elements, each of which is fed with only a single narrowband signal, and connected via a switching network, as shown in Fig.~\ref{fig_IM_ISAC_antennaFrequency}(a). Let $f_{n,p}$ denote the frequency of the waveform fed to the $n$-th antenna at the $p$-th pulse with the PRI of $T$. Then, the transmitted signal at the $n$-th antenna and the $p$-th pulse is 
	\begin{align}
		x_n(t,p) = w_n(\theta_r,f_{n,p}) \mathrm{rect}\left(\frac{t-p T}{T}\right)e^{\mathrm{j} 2\pi f_{n,p}t},
	\end{align}
	where $w_n(\theta_r,f_{n,p}) = e^{\mathrm{j} 2\pi (n-1)f_{n,p}\frac{{d}}{c_0} \sin \theta_r  }$ represents the array response for the target direction $\theta_r$. Now, the $N\times 1$ transmitted signal is written as 
	\begin{align}
		\mathbf{x}(t,p) = \mathbf{w}(\theta_r,p)   \mathrm{rect}\left(\frac{t-p T}{T}\right)e^{\mathrm{j} 2\pi f_{n,p}t},
		\label{transmittedSignal1}
	\end{align}
	where $\mathbf{w}(\theta_r,p)   = \left[w_1(\theta_r,f_{1,p}), \cdots, w_N(\theta_r,f_{N,p})  \right]^\textsf{T}\in \mathbb{C}^{N}$. Assume that each of the $N$ antennas are fed with monotone waveforms of distinct frequencies. Then, $N$ out of $K$ ($K>N$) frequencies are used. Thus, the number of possible frequency permutations is $C(N,K)$, which yields that $\lfloor N \mathrm{log}_2 K \rfloor$ information bits can be conveyed via IM~\cite{im_majorcom_IndexMod_Huang2020May,spim_JRC_FRAC_IM_Ma2021Oct}.

	Without reducing the array size, partitioning the array into multiple groups also gives the opportunity to perform IM. Suppose that the antenna array is composed of $G$ group of antennas, as shown in Fig.~\ref{fig_IM_ISAC_antennaFrequency}(b). Now, define 
	$f_{n,p,g}$ as the waveform frequency corresponding to the $n$-th antenna in the $g$-th group, where $g = 1,\cdots, G$. Furthermore, we can assume that the waveform frequencies are selected from the set of $K$ frequencies, i.e., $\{\tilde{f}_1,\cdots, \tilde{f}_K \}$.  Then, the subarrays of size $\frac{N}{G}$ are used to transmit waveforms of $G$ out of $K$ distinct frequencies. Since all the array elements are employed, the selection matrix is defined as an $N\times N$ diagonal matrix $\tilde{\mathbf{Q}}(p,g)\in \{0,1\}^{N\times N}$, where $[\tilde{\mathbf{Q}}(p,g)]_{n,n} = 1$ if the $n$-th antenna is used in the $g$-th group to transmit $f_{n,p,g}$, and 0 otherwise. Then, the $N\times 1$ transmitted signal is given by 
	\begin{align}
		\tilde{\mathbf{x}}(t,p) = \sum_{g = 1}^{G}\tilde{\mathbf{Q}}(p,g) \mathbf{w}(\theta_r,g,p)   \mathrm{rect}\left(\frac{t-p T_0}{T_0}\right)e^{\mathrm{j} 2\pi f_{n,p,g}t},
	\end{align}
	where $\mathbf{w}(\theta_r,g,p) = \left[w_1(\theta_r,f_{1,p,g}), \cdots, w_N(\theta_r,f_{N,p,g})  \right]^\textsf{T}\in \mathbb{C}^{N}$. Compared to the IM scheme in Fig.~\ref{fig_IM_ISAC_antennaFrequency}(a), the grouped subarray technique in Fig.~\ref{fig_IM_ISAC_antennaFrequency}(b) provides  $C(G,K) \frac{N!}{(N/G)^G}$ allocation patterns. Therefore, the maximum number of information bits that can be conveyed approximately is $\lfloor G\log_2 K \rfloor + \lfloor N\log_2 G \rfloor$~\cite{im_majorcom_IndexMod_Huang2020May}.

	\subsection{Joint Antenna, Frequency and Phase Index Modulation}
\begin{figure*}
		\centering
		{\includegraphics[draft=false,width=.45\textwidth]{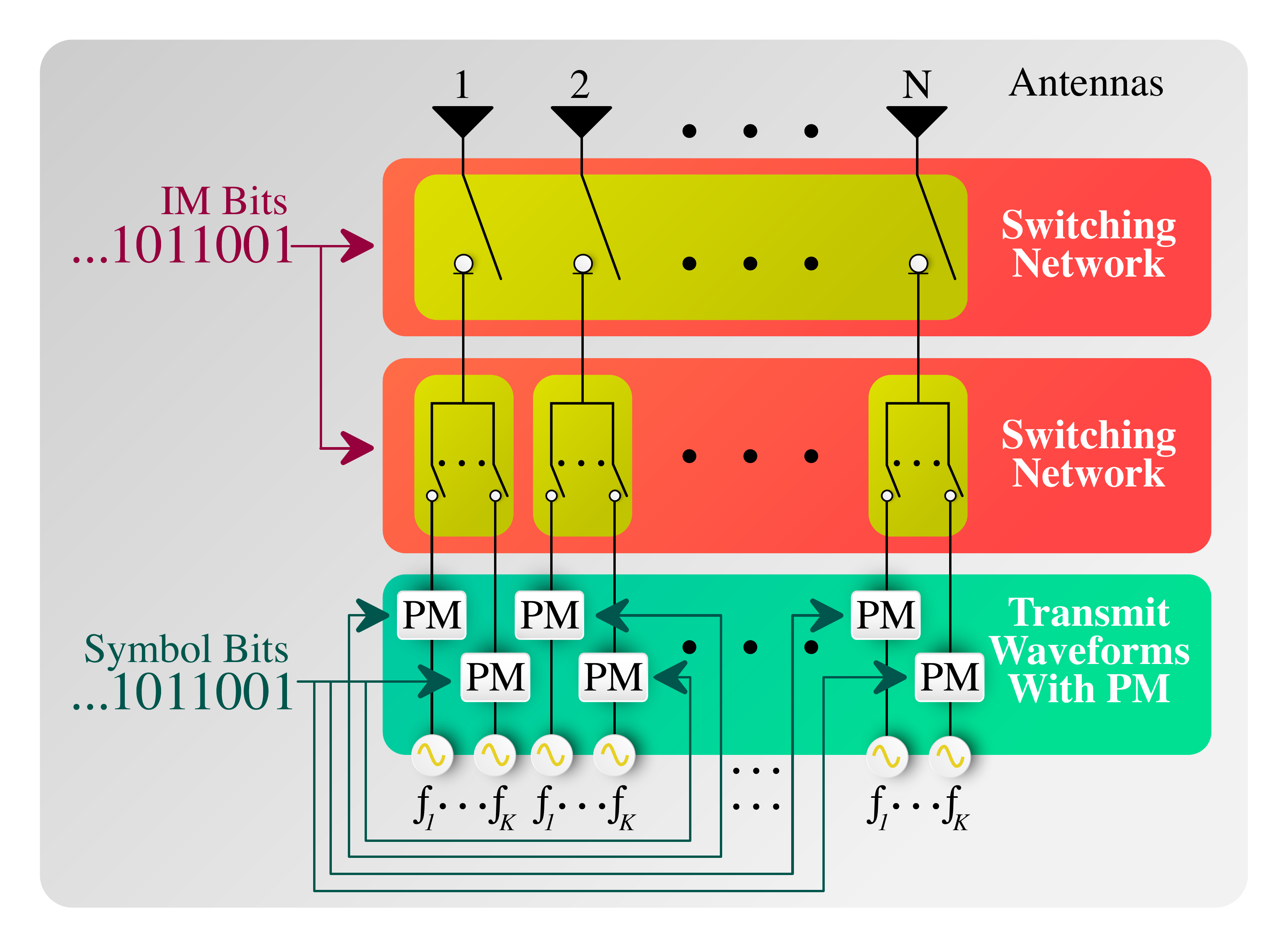} } 
		\caption{Joint antenna, frequency and phase modulation for IM-ISAC. The data stream is split into two portions. While IM bits are utilized via selection of $N_s$ out of $N$ antennas and $N_s$ out of $K$ waveform frequencies, the remaining bits are phase-modulated via $M$-ary constellation.     }
		\label{fig_IM_ISAC_antennaFrequencyPhase}
	\end{figure*}

	Instead of solely employing IM in ISAC~\cite{im_ISAC_sparseArray_Wang2018Aug,im_majorcom_IndexMod_Huang2020May}, the modulation of the symbol bit further improves the communication rate. Motivated by this, an FMCW-based IM-ISAC technique is developed for vehicular applications ~\cite{spim_JRC_FRAC_IM_Ma2021Oct}. Specifically, the transmitter employs FMCW signaling via PM to convey the communication messages. To enhance the IM capability, a subset of the antennas are activated during each pulse with different frequencies, as shown in Fig.~\ref{fig_IM_ISAC_antennaFrequencyPhase}. Define $s(t)\in \mathbb{C}$ as a baseband FMCW waveform as
	\begin{align}
		s(t) = \mathrm{rect}\left( \frac{t}{T} \right) e^{\mathrm{j}\kappa \pi t^2   }, \hspace{10pt} 0\leq t \leq \tilde{T},
	\end{align}
	where $\kappa$ is the frequency modulation rate of FMCW with the duration of $\tilde{T}$, and we have $T \leq  \tilde{T}$ for the PRI $T$. Define $\varphi_{n,p}$ as the phase modulated on the FMCW signal in the $p$-th radar pulse at the $n$-th antenna element. Then, the transmitted waveform at the $n$-th antenna is given by
	\begin{align}
		\bar{x}_n(t,p) = s(t - p T) e^{\mathrm{j}2\pi f_{n,p} t } e^{\mathrm{j} \varphi_{n,p}  }, \label{frac_x}
	\end{align}
	where $f_{n,p} = f_c + k_{n,p} \Delta f$ denotes the selected carrier frequency from the set $\{f_c + k_{n,p} \Delta f | k_{n,p} = 0,\cdots, K-1  \}$, which includes $K$ distinct frequencies with a frequency step of $\Delta f = \frac{ 1}{\kappa T}$. In order to embed the information bits, $N_s$ distinct communication bits are first modulated via PM with $M$-ary constellation. Thus, the number of bits conveyed via PM is $N_s \log_2 M$. Then, $N_s$ carrier frequencies are selected from $K$ possible frequencies, yielding $\lfloor \log_2 C(N_s,K) \rfloor $ selection patterns. Finally, the phase- and frequency-index modulated waveforms are assigned to $N_s$ out of $N$ antennas to convey $N_s!$ different symbols in the permutation of the waveforms. As a result, the total number of transmitted bits is $ {{N_\mathrm{S}\log_2 M} } + \lfloor \log_2 C(N_s,K)\rfloor + \lfloor \log_2 C(N_s,N) \rfloor + \lfloor \log_2 N_s! \rfloor $.

	Although the FMCW-based approach in~\cite{spim_JRC_FRAC_IM_Ma2021Oct} provides enhanced communication rate, it does not fully exploit the DoF of all domains (e.g., antenna, frequency and phase). In order to fully leverage the diversities of the integrated waveform, an FH-based signaling strategy is developed for IM-ISAC in~\cite{im_hybrid_Xu2022Nov}. In particular, a high dimensional variable space is created over frequency code, initial phase and transmit antenna index to embed the communication information. 	In FH, the pulse interval $T$ is divided into $H$ subpulses, which are also referred to as hops~\cite{fh_MIMO_radar_Wu2021Dec}. Then, a sinusoidal signal is transmitted in each hop.   Define $c_{n,h}$ as the FH code selected from the set of $K$ frequencies, i.e., $\{c_1,\cdots, c_K \}$  at the $h$-th subpulse transmitted at the $n$-th antenna. Then, the transmitted FH waveform is given by
	\begin{align}
		u_{n}(t,h) = e^{\mathrm{j}2 \pi c_{n,h} \delta_f t  } v(t - h \delta_t), 
	\end{align}
	where $ \delta_f$ and $\delta_t = T/H$ represent the unit frequency step and the duration of each subpulse, respectively. The rectangular pulse $v(t)$ is defined as $	v(t) = \left\{	\begin{array} {cc}
		1, & 0\leq t \leq \delta_t, \\
		0, & \mathrm{otherwise}
	\end{array} \right.$. 	Then, the transmitted waveform at the $h$-th subpulse and the $n$-th antenna can be represented as
	\begin{align}
		z_{n}(t,h) = e^{\mathrm{j}\vartheta_{n,h}   }u_n(t,h) ,
	\end{align}
	where $\vartheta_{n,h}$ is the phase term, which is modulated corresponding to FH code $c_{n,h}$, and it is formulated as $		\vartheta_{n,h} = \vartheta_{n,h,{i}} + \zeta_{n,h}$, 	where $\vartheta_{n,h,{i}}$ denotes the initial phase represented by the binary phase-shift-keying (BPSK) symbol as $	\vartheta_{n,h,{i}} = \left\{	\begin{array} {cc}
		\pi, & {i} = 1, \\
		0, & {i} = 0.
	\end{array} \right.$, and $\zeta_{n,h}$ is the phase associated with the antenna index $n$ as  $\zeta_{n,h} = \frac{(n-1)\pi}{N} - \angle\{[\mathbf{a}(\theta_r)]_n  \}  $, where $[\mathbf{a}(\theta_r)]_n$ denotes the $n$-th entry of the transmit steering vector corresponding to the $r$-th target. In this FH-IM-ISAC scheme, the cardinality of the hybrid dictionary over frequency, phase and antenna indices can be as large as $C(K,N) \cdot N! \cdot 2^{N}$, which yields the communication rate of $H \cdot  (N $ $+ \lfloor \log_2 (C(K,N)\cdot N!) \rfloor  )$~{\cite{im_hybrid_Xu2022Nov}}.


	%
	%
	
	%
	%

	%

	%

	



	
	\subsection{Spatial Path Index Modulation}
	Majority of the IM schemes presented above~\cite{im_ISAC_sparseArray_Wang2018Aug,im_majorcom_IndexMod_Huang2020May,spim_JRC_FRAC_IM_Ma2021Oct,jrc_spim_sm_Ma2021Feb,jrc_generalized_SM_Xu2020Sep} suffer from low communication rates since the information embedding is usually applied in slow-time over radar pulses, thereby bounding the communication rate by the PRF. Besides, a communication-centric approach is developed in~\cite{elbir_SPIM_MMWAVE_RadarConf_Elbir2022Nov,elbir_SPIM_ISAC_THZ_Elbir2023Mar}, wherein the transmitter (i.e., the BS) employs hybrid analog/digital beamformers generating simultaneous beams toward both radar targets and communication users. To further improve the communication rate, the diversity of spatial paths between the BS and communication user is exploited for SPIM. Assume there are $L = L_R + L_C$ spatial paths received from radar targets $(L_R)$ and the communication user $(L_C)$. Then, a switching network is employed to perform IM by connecting $N_\mathrm{RF} = L_s + L_R$ RF chains to $L$ taps of the analog beamformer to exploit $L_R$ paths for the radar targets and $L_s$ out of $L_C$ paths for the communication user via SPIM. As a result, the number of additional communication bits via SPIM is $\lfloor \log_2 C(L_s, L_C) \rfloor$. 
	
	Define $\mathbf{F}_\mathrm{RF}^{(i)}\in \mathbb{C}^{N\times N_\mathrm{RF}}$ and $\mathbf{F}_\mathrm{BB}^{(i)}\in \mathbb{C}^{N_\mathrm{RF}\times N_\mathrm{S}}$ as the analog and baseband beamformers for the $i$-th \textit{spatial pattern}, where $N_\mathrm{S}$ is the number of data-streams transmitted via conventional communication system. Then, the $N\times 1$ transmitted signal becomes
	\begin{align}
		\mathbf{x}_\mathrm{SPIM}^{(i)} = \mathbf{F}_\mathrm{RF}^{(i)}\mathbf{F}_\mathrm{BB}^{(i)}\mathbf{s},
	\end{align}
	where $\mathbf{s}\in \mathbb{C}^{N_\mathrm{S}}$ includes the transmitted data-streams. The structure of the analog beamformer is given by $		\mathbf{F}_\mathrm{RF}^{(i)} = \left[ 	\mathbf{F}_\mathrm{R} \hspace{2pt}  | \hspace{2pt}	\mathbf{F}_\mathrm{C}^{(i)}\right],$
	where $	\mathbf{F}_\mathrm{R} = \left[\mathbf{a}(\theta_1),\cdots, \mathbf{a}(\theta_{L_R})  \right] \in \mathbb{C}^{N\times L_R}$ is the radar-only beamformer corresponding to $L_R $ radar target paths whereas $		\mathbf{F}_\mathrm{C}^{(i)}  \in \mathbb{C}^{N\times L_s}$ represents the communication-only beamformer corresponding to the path directions from the user, and it is defined as $		\mathbf{F}_\mathrm{C}^{(i)} = \left[\mathbf{a}(\phi_1),\cdots, \mathbf{a}(\phi_{L_C})  \right]\mathbf{B}^{(i)}, $	where $\mathbf{B}^{(i)}$ is an $L_s\times L_C$ selection matrix, selecting the $L_s$ steering vector from $L_C$ spatial paths for the $i$-th spatial pattern as {$\mathbf{B}^{(i)} = \left[\mathbf{b}_{i_1},\cdots, \mathbf{b}_{i_{L_C}}  \right]$}, where $\mathbf{b}_{i_l}$ is the $i_{l}$-th column of identity matrix $\mathbf{I}_{L_s}$. The analog and digital beamformers are then optimized by maximizing the SE of the SPIM-ISAC system, which is defined as~\cite{elbir_SPIM_ISAC_THZ_Elbir2023Mar}
	\begin{align}
		\mathrm{SE}_\mathrm{SPIM-ISAC} = \log_2   &\left( \frac{2^{S} }{(2\sigma_n^2)^{\bar{N}}  } \right)  \nonumber \\
		 -
		\frac{1}{S} \sum_{i = 1}^{S}& \log_2 \left( \sum_{j = 1}^{S} \mathrm{det}\{\boldsymbol{\Sigma}_i  + \boldsymbol{\Sigma}_j \}^{-1}  \right),
	\end{align}
	where {$S =  2^{\lfloor \log_2 C(L_s,L_C)\rfloor}$  denotes the number spatial patterns}, and $\sigma_n^2$ is the noise variance and $\boldsymbol{\Sigma}_i\in \mathbb{C}^{\bar{N}\times \bar{N}}$ is defined as $	\boldsymbol{\Sigma}_i = \mathbf{I}_{\bar{N}} + \frac{1}{\sigma_n^2 N_\mathrm{S}} \mathbf{H} \mathbf{F}_\mathrm{RF}^{(i)}\mathbf{F}_\mathrm{BB}^{(i)} \mathbf{F}_\mathrm{BB}^{(i)^\textsf{H}}\mathbf{F}_\mathrm{RF}^{(i)^\textsf{H}} \mathbf{H}^\textsf{H} ,$	where  $\mathbf{H}\in \mathbb{C}^{\bar{N}\times N}$ represents the wireless channel matrix. In contrast, the SE of the ISAC without SPIM is defined as $\mathrm{SE}_\mathrm{ISAC} = \log_2 \left(\mathrm{det}\left\{  \boldsymbol{\Sigma}_1  \right\}   \right)   $, where $i=1$ denotes the spatial pattern which includes only the strongest path between the BS and the user.

	\begin{figure*}[t]
		\centering
		\subfloat[]	{\includegraphics[draft=false,width=\columnwidth]{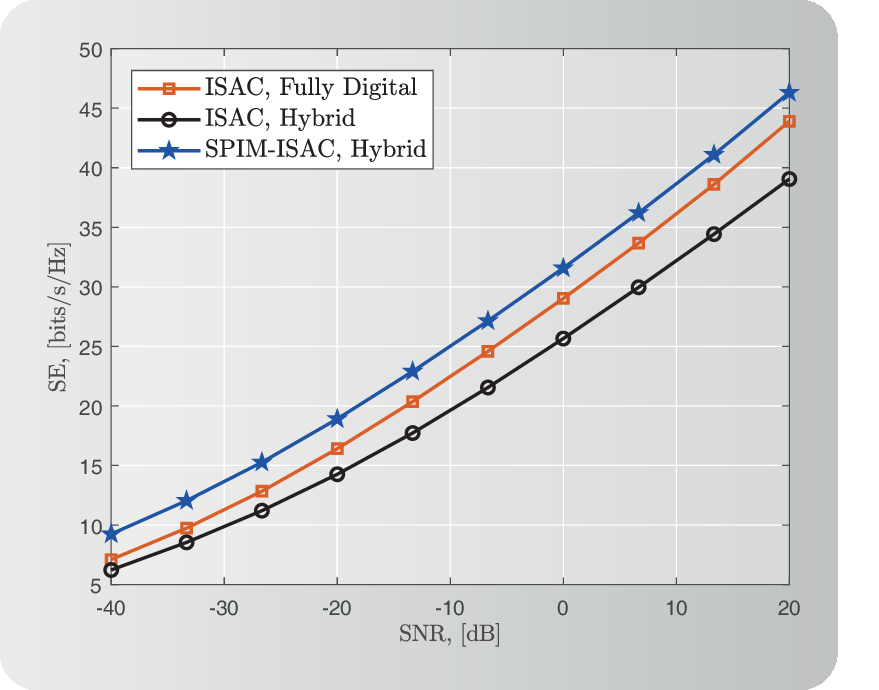} } 
		\subfloat[]	{\includegraphics[draft=false,width=\columnwidth]{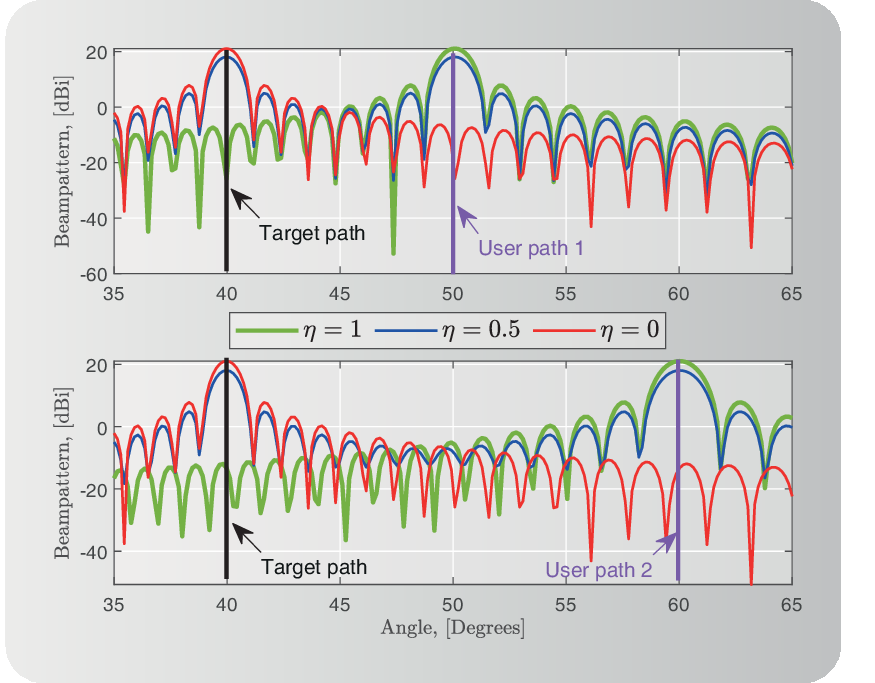} } 
		\caption{(a) SE versus SNR for the radar-communications trade-off parameter $\eta = 0.5$. (b) Beampattern for $R=1$, $(L_C,L_s) = (2,1)$ when $(\theta_1, \phi_1)$ is (top) $(40^\circ, 50^\circ)$ ($i=1$) and (bottom) $(40^\circ,60^\circ)$ $(i=2)$, respectively, for $\eta = \{0,0.5,1\}$. 
		}
		
		\label{fig_SPIM}
	\end{figure*}

	Fig.~\ref{fig_SPIM} presents the performance of the SPIM-based hybrid beamforming for ISAC in terms of signal-to-noise ratio (SNR)~\cite{elbir_SPIM_ISAC_THZ_Elbir2023Mar}. We observe from Fig.~\ref{fig_SPIM}(a) that SPIM-ISAC achieves a significant improvement compared to conventional ISAC with both hybrid and fully digital beamforming thanks to additional transmitted information bits via SPIM. Fig.~\ref{fig_SPIM}(b) shows the beampattern of the SPIM-ISAC hybrid beamformer when only $S = 2$ ($i \in \{1,2\}$) spatial patterns are used. In this scenario, $R=1$ and $(L_C,{L_s}) = (2,1)$. The target is located at  $\theta_1 = 40^\circ$ while the user paths are received from $\phi_1 = 50^\circ$ ($i =2$) and $\phi_2 = 60^\circ$ ($i =2$), respectively. We see that the beampattern is suppressed at the target direction when the radar-communications trade-off parameter $\eta \rightarrow 1$. Conversely, the beampattern at the user locations is minimized when $\eta \rightarrow 0$.
	

	\section{Challenges in IM-ISAC}
	The implementation of IM-ISAC is currently under extensive investigation by the ISAC community. While there exist a few experimental works for hardware prototypes~\cite{jrc_spim_sm_Ma2021Feb,prototype_ISAC_R2022Nov}, the design of IM-ISAC poses critical challenges. Herein, we state some of the leading research challenges in waveform design, latency as well as the receiver design in IM-ISAC.
	
	\subsection{Waveform Design}
	Aside from the implementation challenges, waveform design constitutes a major hardship in ISAC as the current research focuses on embedding communication bits into the radar waveforms. The design of IM-ISAC becomes even more challenging as additional information bits are also taken into account over multiple resource entities~\cite{waveforDesign_SpectralMod1_Yu2022Apr}.  While OFDM signaling has been a favorable waveform for communications, its usage in IM-ISAC puts demands on power consumption, synchronization between the transmitter and the receiver as well as receiver design to demodulate the IM bits~\cite{im_ofdm_ISAC_Li2024Feb}. Rather than OFDM, FMCW waveform is commonly used in radar applications, especially in automotive radars, due to its simplicity and low complexity. To make it more communication-friendly, FMCW-based IM-ISAC designs are introduced in~\cite{spim_JRC_FRAC_IM_Ma2021Oct}. While achieving enhanced multiplexing of the communication bits in the radar signal over multiple resource entities, they still suffer from low SE due to the employment of chirp-like signals~\cite{jrc_spim_sm_Ma2021Feb}. Alternatively, FH enjoys high data rate by employing IM over multiple resource entities as well as  transmitting the information bits over multiple radar subpulses~\cite{fh_MIMO_radar_Wu2021Dec,im_FH_Eedara2022Jan}. Nevertheless, this improvement in the communication data rate requires a proper demodulation architecture at the receiver. Furthermore, synchronization of the radar subpulses is another challenging issue as the inter-hop-interference (IHI) is introduced to the receiver. As a result, each of the possible state-of-the-art IM-ISAC waveforms has certain trade-offs in various aspects, e.g., hardware, communication rate, sensing resolution as well as complexity, thereby suggesting that there is still room to strike a balance among design criteria.

	\subsection{Latency}
	Most of the IM-ISAC designs work by embedding the information bits over one or multiple radar pulses on the slow-time scale. Thus, the symbol rate is limited by the radar PRF, which is not capable of carrying data at the desirable rates for commercial communication services, and also cause clutter modulations from the radar perspective~\cite{im_FH_Eedara2022Jan}. Promising solutions include FH-IH signaling~\cite{im_hybrid_Xu2022Nov,fh_MIMO_radar_Wu2021Dec}, SPIM~\cite{elbir_SPIM_MMWAVE_RadarConf_Elbir2022Nov,elbir_SPIM_ISAC_THZ_Elbir2023Mar} or CSK-based solutions~\cite{im_FH_Eedara2022Jan} to break this limitation and increase the communication rate. Another approach, similar to SPIM, DAM is proposed in~\cite{im_delayAlignmentMod_Xiao2023Apr} to account for path-based beamforming such that each of the multipath components transmitted from the ISAC transmitter arrives at the receiver simultaneously, thereby increasing the data rate.  Although this approach does not involve IM, it provides higher communication rates compared to OFDM signaling.
	In essence, communication-centric IM-ISAC architectures are of interest as higher communication rate can be achieved as compared to the radar-centric designs. In fact, this leads to the scenario that the IM-ISAC architectures will closely resemble the conventional communication architectures, which suggest sensing-aided communication paradigms~\cite{elbir_thz_jrc_Magazine_Elbir2022Aug}.
	
	\subsection{Receiver Design}

	The receiver should take into account the demodulation {of} the IM bits. In OFDM signaling with IM over subcarrier indices, this is done by simply checking the (de)activated subcarriers via applying orthogonal waveforms~\cite{im_ofdm_ISAC_Li2024Feb}. Matched filtering is performed to recover the sparse array data when the IM bits are modulated over antenna indices~\cite{im_antennas_generalized_Zhang2013Oct}. As the IM strategy involves two or more transmission media (e.g., antenna and frequency), the complexity of signal processing at the receiver becomes high in seeking to demodulate the IM bits over multiple domains. Apart from communication receiver design, the radar receiver should take into account the interferences such as sampling jitter and phase noise during collecting coherent data samples to ensure synchronization and stability. {In the following, we discuss the recovery of the IM bits for different IM-ISAC schemes. }
	
	\begin{itemize}
		\item {\textit{Subcarrier Index Modulation:} The communication and radar symbols are distributed over the subcarriers while a portion of the subcarriers are intentionally left empty. Thus, the estimation of these null subcarriers at the receiver yields the location of radar and communication symbols~\cite{spim_OFDM_IM_JRC_Sahin2021}. Consider the transmitted signal given in (\ref{si_xs}), and define $\mathbf{\overline{\mathbf{Y}}}\in \mathbb{C}^{B \times  K}$ as the frequency domain received signal for the transmission of $B$ symbols over $K$ subcarriers, $K_s$ of which are empty. Then, the location of the null subcarriers in each subblock is found via solving
			\begin{align}
				k^\star_b = \arg \minimize_k | [\overline{\mathbf{Y}}]_{b,k}  |^2,
			\end{align}
			which is then used to recover the communication symbols according to a look-up table, which is known to the receiver.      	}

		\item {\textit{Antenna Index Modulation:} At the communication receiver, the IM bits are recovered via employing a dictionary of sparse array configurations~\cite{im_ISAC_sparseArray_Wang2018Aug}. Assuming synchronization between the communication receiver and the ISAC transmitter, the IM symbols can be recovered by minimizing the the Euclidean distance between the normalized received data at the $p$-th pulse in (\ref{ai_yc}), i.e., $\frac{1}{\alpha_C}\mathbf{y}_C(p)$, and the candidate $N_s\times 1$ sparse steering vector $\bar{\mathbf{a}}_i(\phi,p) $ as
			\begin{align}
				{	\minimize_i 	\|  \frac{1}{\alpha_C}\mathbf{y}_C (p) -   \bar{\mathbf{a}}_i(\phi,p)   \|_2, }
			\end{align}
			where $\bar{\mathbf{a}}_i(\phi,p)  = \boldsymbol{\Pi}_i \mathbf{a}(\phi)  $, and $\boldsymbol{\Pi}_i\in \mathbb{C}^{N\times N}$ is a selection matrix comprised of $\{0,1\}$	 corresponding to the $i$-th candidate sparse array configurations for $i = 1,\cdots, C(N_s, N)$.  The number of communication bits that can be transmitted per symbol is then given by $\lfloor \log_2 C(N_s,N)\rfloor$.  }

		In a similar way, the MIMO radar receiver performs match filtering for the received target echo signal in {(\ref{ai_ytp}), i.e., $\mathbf{y}(t, p) $,} with $\boldsymbol{\Psi}(t)$, and obtains the $\bar{N}N_s\times 1$ signal $\mathbf{y}_{R}(p)$ as
		\begin{align}
			\mathbf{y}_{R}(p) & = \mathrm{vec} \left\{  \int_{T} \mathbf{y}(t,p) \boldsymbol{\Psi}^\textsf{H}(t) dt \right\}
		\nonumber \\
		&	=  \sum_{r = 1}^R \alpha_r (p) [\mathbf{Q}(p) \mathbf{a}(\theta_r)]  \otimes \tilde{\mathbf{a}}(\theta_r)  + \mathbf{n} (p),
		\end{align}
		where {$\mathbf{n} (p) =  \mathrm{vec} \left\{ \int_{T} \mathbf{n} (t,p) \boldsymbol{\Psi}^\textsf{H}(t)dt \right\} \in \mathbb{C}^{\bar{N}N_s}$} is the additive noise term after  matched filtering.

		\item {\textit{Joint Antenna, Frequency and Phase Index Modulation:} At the user end, the $p$-th received signal after collecting $U$ samples, $\widetilde{\mathbf{y}}_C(p)\in \mathbb{C}^{U}$ can be written as 
			\begin{align}
				\widetilde{\mathbf{y}}_C(p) = \boldsymbol{\Phi}\boldsymbol{\xi}(p) + \widetilde{\mathbf{n}}(p),
			\end{align}
			where $\boldsymbol{\xi}\in \mathbb{C}^{NK}$ is a block-sparse vector including the transmitted IM symbols at the $p$-th pulse, and $\widetilde{\mathbf{n}}\in \mathbb{C}^{U}$ represents the noise term. $\boldsymbol{\Phi}$ is  an $U\times NK$ dictionary matrix comprised of the set of $U\times N$ matrices composed of the channel response vectors~\cite{spim_JRC_FRAC_IM_Ma2021Oct}. Then, the IM symbols can be recovered via
			\begin{align}
				\hat{\boldsymbol{\xi}}(p) = \arg \minimize_{	{\boldsymbol{\xi}}(p)} \| 	\widetilde{\mathbf{y}}_C(p) -    \boldsymbol{\Phi}\boldsymbol{\xi}(p)  \|_2^2.
			\end{align}

		}

	\end{itemize}

	{\subsection{Sensing Performance}
		Maintaining the sensing performance in IM-ISAC is also a challenging issue as a portion of the RF components (e.g., subcarriers, time-slots and antennas) are shared or dedicated to communication task. This may cause a loss in terms of several performance metrics for radar sensing, e.g., SNR, CRB, radar ambiguity function as well as the beampattern sidelobe level. For example, in antenna index modulation, a sparse array is employed to generate IM bits via the indices of the (in)active antennas, which reduces the SNR for radar sensing~\cite{im_ISAC_sparseArray_Wang2018Aug} as well as degrading the CRB level for target parameter estimation~\cite{im_constellationExtension_Memisoglu2023May}. The 
		incidental selection of the antenna elements in the sparse array also causes undesired beampattern with high sidelobe levels. Therefore, one cannot employ all the subarray configurations for IM. This poses a trade-off for the communication rate with less subarray configurations and providing desired beampattern design for radar sensing. To this end, instead of all the subarray configurations, only a subset which satisfies the criterion of desired sidelobe level can be selected~\cite{im_ISAC_sparseArray_Wang2018Aug}.  In order to further improve the sensing performance, one possible approach is to exploit the full array to improve the sensing performance while still performing IM is devised in~\cite{im_majorcom_IndexMod_Huang2020May}. Specifically, the whole array is divided into groups, each of which are fed with waveforms of the distinct frequencies for IM. Nevertheless, the MAJoRCom approach in~\cite{im_majorcom_IndexMod_Huang2020May} exhibits a limited sensing performance due to poor radar ambiguity function, especially for resolving the targets with the same range and velocities. Further enhanced sensing performance is achieved by the joint antenna, frequency and phase modulation technique based on FMCW radar~\cite{spim_JRC_FRAC_IM_Ma2021Oct}, and it is shown that close to wideband MIMO radar performance is achieved while performing IM.

	}


	\section{Summary}
	\label{sec:Summary}
	The plethora of IM schemes, their possible variants, and their respective benefits present a swiss-knife procedure for selecting the most efficient and appropriate IM scheme for  a specific ISAC application. This article presented an overview of major IM-ISAC techniques,  and delineated their transmitter architectures, implementation of IM over various resource entities, and their achievable communication data rates as well as the corresponding radar sensing capabilities.
	
	Various waveform designs are presented to embed the information bits into the radar waveforms. The communication rate improves as the number of IM domains increases. This, however, entails complexity involved at both transmitter to perform IM as well as at the receiver to demodulate the IM bits. While OFDM-IM has a simple architecture, limited IM bits can be conveyed. Similarly, IM over sparse array design causes distortions in the beampattern and array gain loss. While performing IM, maintaining the full usage of the IM resources, especially the subcarriers and antennas, is of great importance as it determines the radar sensing and detection performance. The radar-centric architectures, e.g., FMCW, suffer from the low communication data rates limited by the radar PRF. In contrast, the communication-centric designs, e.g., FH-ISAC and SPIM-ISAC, provide higher rates via transmitting more IM bits while providing satisfactory radar performance. 
	
	Implementing practical IM-ISAC design poses new challenges to be tackled. This introduces new opportunities to develop efficient IM strategies as well as new applications. In particular, future work includes orthogonal time frequency space (OTFS) modulation with IM, which is well investigated in communications~\cite{otfs_IM_comm_Qian2023Jun}, can be considered for ISAC. Furthermore, IM for intelligent reflecting surface (IRS) aided ISAC~\cite{elbir_IRS_ISAC_Elbir2022Apr} opens up new applications, especially in vehicular scenarios. Besides, data-driven techniques, such as machine learning (ML)~\cite{elbir2021ICASSP_FL_HB_SPIM,elbir_sparseArrayRadar_chapter_CSDOAEst}, have shown effectiveness for easing complexity in data-rich applications, therefore, can be leveraged in demodulation of IM signals, physical layer design~\cite{elbir2022Nov_Beamforming_SPM} and parameter estimation~\cite{elbir2021JointRadarComm,elbir_SPIM_MMWAVE_RadarConf_Elbir2022Nov}.



	
	\balance
	\bibliographystyle{IEEEtran}
	\bibliography{references_128}

\end{document}